\documentclass[10pt]{elsart}
\usepackage{dcolumn}
\usepackage{graphicx}
\usepackage{eso-pic}
\usepackage{threeparttable}
\usepackage{cite}
\usepackage{rotating}
\usepackage{xcolor,colortbl}
\usepackage{amsmath}

\begin{document}

\begin{frontmatter}
\title{TENDL-astro: a new nuclear data set for astrophysics interest}
\author[psi]{D. Rochman}, \author[iaea]{A. Koning}, \author[ulb]{S. Goriely} and \author[cea,saclay]{S. Hilaire}
\address[psi]{Reactor Physics and Thermal hydraulic Laboratory, Paul Scherrer Institut, Villigen, Switzerland}
\address[iaea]{Nuclear Data Section, IAEA, Wagrammerstrasse 5, 1400, Vienna, Austria}
\address[ulb]{Institut d'Astronomie et d'Astrophysique, Universit\'e Libre de Bruxelles, Campus de la Plaine, CP-226, 1050, Brussels, Belgium}
\address[cea]{CEA, DAM DIF, F-91297 Arpajon, France }
\address[saclay]{Universit\'e Paris Saclay, CEA, LMCE, F-91680 Bruy\`eres-Le-Ch\^atel, France}

\begin{abstract}
In this work, we are presenting a new database of astrophysical interest, based on calculations performed with
the nuclear reaction code TALYS. Four quantities are systematically calculated for over 8000 nuclides: cross
sections, reaction rates, Maxwellian Averaged Cross Sections (or MACS) at 30 keV and partition functions. For cross sections
and reaction rates, nine reactions are considered, induced by neutron, proton or alpha. The main complement of this database compared to existing ones is that the impact of reaction models ({\it e.g.} level density, gamma strength function, and optical model) is estimated by varying 9 different models, and by proposing calculated values for each of them, together with averages, standard deviations and other statistical quantities. This new database, called TENDL-astro, version 2023, is available online (https://tendl.web.psi.ch/tendl\_2023/astro/astro.html) and linked to the well-known TENDL database, used in a variety of applications.
\end{abstract}


\end{frontmatter}

\section{Introduction}
Nuclear data are undeniably key quantities for applications based on nuclear reactions. In the present context, they principally represent probabilities of interaction, or cross sections 
(induced by neutron or charged particles) and eventually derived quantities ({\it e.g.} reaction rates). Their accurate knowledge affects a very broad range of applications~\cite{PhysRevResearch.4.021001}, 
from energy production, waste management and medical applications to the evolution of stars and stellar nucleosynthesis. Sensible efforts are dedicated to improve their predictions based on modelling and when possible, 
measurements. Naturally, different applications involve different ranges of interest, regarding nuclides (with various orders of magnitude for their  half-lives), incident particles (neutrons, protons, 
and others), incident energy ranges, and reactions.\\
In astrophysics, a  number of publications show the importance of nuclear data for a large number of nuclides, various reactions and incident particles, as well as energy 
ranges~\cite{Arnould20,fspas.2023.1243615,Kankainen2022,Langanke2023}. This is justifying experimental efforts~\cite{universe8050255}, as well as the application of a variety of calculations (being from 
general-purpose  nuclear data libraries~\cite{plompen2020}, or from dedicated databases~\cite{bruslib,Cyburt_2010}). We propose to systematically estimate a number of cross 
sections and other quantities (see section~\ref{interest}), based on nuclear reaction models. Such models range from a pragmatic phenomenological approach to purely microscopic calculations, thus 
proposing a large number of estimated quantities due to model variations (see section~\ref{calculations}). They are included in the TALYS 1.96 nuclear reaction code~\cite{talys} corresponding 
the version used in this work, and results of these calculations are presented in the online TENDL-2023 library, with a number of comparisons in section~\ref{verif}. \\
In TALYS, there are about 20 different models which can influence a specific cross section such as for neutron capture, and the user can avoid 
explicitly selecting any of them as a default selection is also proposed. Thus, the user does not systematically recognize that such a 
choice can impact calculated quantities.
The result is nevertheless potentially dependent on the hidden or definite model selection. We are therefore proposing to make this 
choice visible, and to modify 
it, quantifying the impact on a number of quantities of interest for astrophysics.   \\
This new proposed database, called ``TENDL-astro'', version 2023, aims at complementing the existing ones, by providing results from up-to-date models, and also by providing estimated uncertainties 
due to model variations. It is based on the large accumulated experience in the TALYS code, but also in the production of the TENDL libraries since 2008~\cite{KONING20191}, its application to MACS 
calculations and uncertainties~\cite{ROCHMAN2017109,ROCHMAN2010669}. Details of the database production and results are presented in the following sections.

\section{Quantities of interest}\label{interest}
Many quantities of interest can be defined for astrophysical purposes. The present database will be limited to some of them, calculated with TALYS: cross sections (for incident projectile energies ranging between 0 and 20 MeV), reaction rates  and normalized partition function $G(T)$ (for temperatures from 10$^5$ to 10$^{10}$ K), and the laboratory Maxwellian Averaged (n,$\gamma$) Cross Sections (or MACS) at 30 keV. General descriptions are given below. 

\subsection{Cross sections}\label{crossSections}
 For the  cross sections of interest in this work, the compound nucleus approximation described by the Hauser-Feshbach formalism is used in 
TALYS for all incident particle energies~\cite{talys,goriely2008}. 
It is well known that when the number of available states in the compound nucleus is relatively small, the capture reaction is known to 
be possibly dominated by direct electromagnetic transitions to a bound final state rather than through a compound nucleus intermediary, 
as described by the Hauser-Feshbach formalism. This direct capture mechanism can be satisfactorily described with the perturbative 
approach known as the potential model \cite{Xu14}. It is now well accepted that the direct capture is important and often dominant at the very low 
energies of astrophysical interest for light or exotic nuclei systems for which few, or even no resonant states are available. However, 
significant uncertainties still affect the direct capture predictions. These are related to the determination of the nuclear structure 
ingredients of relevance, i.e the nuclear mass, spectroscopic factor, neutron-nucleus interaction potential and excited level scheme. 
Due to such uncertainties, the direct capture contribution is not included in the present study. 
In addition no individual resonance behavior is calculated at low energy (below a few tens of keV), but the average reaction cross sections are in agreement with either the selected phenomenological or semi-microscopic optical model. A full description of these models can be found in Refs.~\cite{talys,bauge1998,bauge2001,KONING2003231}. \\
 For reactions represented by $I^\mu+a\longrightarrow I'+a'$, the corresponding cross sections are commonly referred to $\sigma_{a,a'}^{\mu}(E)$, where $a$ is the incident particle ({\it e.g.} a neutron), $a'$ is 
the exit channel, $E$ is the incident particle energy, and $\mu$ designates the excited state of the target nucleus $I$ ($\mu=0$ corresponds to the ground state). The full equation of $\sigma_{a,a'}^{\mu}(E)$ in the Hauser-Feshbach formalism used by TALYS can be found in Refs.~\cite{talys,goriely2008} and is not repeated here.\\
Three types of incident particles are considered: neutron, proton and alpha. For each of these particles, three outgoing channels are calculated and presented in the database:
\begin{enumerate}
    \item for neutrons: (n,$\gamma$), (n,p), (n,$\alpha$),
    \item for protons: (p,$\gamma$), (p,n), (p,$\alpha$) and 
    \item for alphas: ($\alpha$,$\gamma$), ($\alpha$,n), ($\alpha$,p).
\end{enumerate}
The incident energy grid spans from 10$^{-5}$ eV up to 20 MeV. 
Note that no  photon-induced reactions are provided, even if a similar method can be applied, since, for astrophysical applications, photodisintegration reaction rates are traditionally derived from the reverse rates on the basis of detailed balance relations \cite{Holmes76}. 

\subsection{Reaction rates and normalized partition function $G(T)$}
The astrophysical reaction rates, for the  reactions (1) to (3) as defined in section~\ref{crossSections}, are obtained following the definition in Ref.~\cite{goriely2008}. It corresponds to the integration of the cross
section from the compound formula $\sigma_{a,a'}^{\mu}$, over a Maxwell-Boltzmann distribution of energies $E$ at a specific temperature $T$ ({\it e.g.} the thermal energy $kT=30$~keV corresponds to a temperature of $T=3.48\times 10^9$ K).\\
Additionally, the target nucleus is not considered in its ground state solely (with $\mu=0$, corresponding to the target ground state spin and an excitation energy $E^0_x=0$), due to the thermodynamic equilibrium in a hot astrophysical plasma (in a stellar interior), and it also exists in excited states. 
In the following, many excited states are now considered, and cross sections $\sigma_{a,a'}^{\mu}(E)$ are used, with $\mu$ corresponding to specific excited states (with a spin $I^\mu$ and an excitation energy $E^\mu_x$).
The distribution of the population of excited states (with $I^\mu$ and $E^\mu_x$) follows a Maxwell-Boltzmann distribution, so that the astrophysical reaction rate $N_A<\sigma\nu>^*_{a,a'}(T)$, also referred here as RR, can be expressed as:

\begin{eqnarray}
    N_A<\sigma\nu>^*_{a,a'}(T) &=& \sqrt{\frac{8}{\pi m}}\frac{N_A}{\sqrt[3]{kT}G(T)}\int_0^\infty \sum_\mu \frac{(2I^\mu +1)}{(2I^0 +1)}\times\nonumber\\
    &&\sigma_{a,a'}^{\mu}(E) \exp\left( -\frac{E+E_x^\mu}{kT} \right)dE ~.  \label{rate}
\end{eqnarray}

The calculated reaction rates (for the 9 reactions defined in the previous section) are obtained with TALYS following Eq.~(\ref{rate}). 
In this equation, $N_A$ is the Avogadro number, $k$ is the Boltzmann constant, $m$ the reduced mass of the incident channel $a$, and $G(T)$ is the temperature-dependent normalized partition function, defined as
\begin{equation}
    G(T)= \sum_\mu \frac{(2I^\mu +1)}{(2I^0 +1)}\times \exp\left( -\frac{E_x^\mu}{kT} \right) ~.
\end{equation}
Both RR and $G(T)$ quantities are provided in the TENDL-astro database, for a set of temperature $T$ ranging between $10^5$ and $10^{10}$~K.

\subsection{Maxwellian Averaged (n,$\gamma$) Cross Sections}
The (n,$\gamma$) MACS of interest corresponds to the one obtained in the laboratory environment, {\it i.e.} with $\mu=0$ (target nucleus in the ground state). This facilitates the comparison between the calculated and measured MACS, for instance from the KADoNiS database~\cite{DILLMANN2014171,dillmann2006}. The expression of the MACS can be derived from
Eq.~(\ref{rate}) by considering $\mu=0$ for the target nucleus:
\begin{eqnarray}
    \text{MACS}~(T) &=& \sqrt{\frac{8}{\pi m}}\frac{N_A}{\sqrt[3]{kT}}\int_0^\infty \sigma_{a,a'}(E) \exp\left( -\frac{E}{kT} \right)dE\label{macs}
\end{eqnarray}
where $\sigma_{a,a'}=\sigma_{a,a'}^{\mu=0}$. 
In the current version of the TENDL-astro database, the MACS values are calculated at $kT=30$~keV.

\section{Considered models for TALYS calculations}\label{calculations}
As presented in the two following subsections, two types of calculations are performed. The first one provides recommended quantities 
(cross sections, reaction rates, MACS and the normalized partition function $G(T)$), based on one set of selected models. The second one presents the variations of these quantities for other set of models, interpreted as ``{\it model uncertainties}''.

\subsection{Nominal calculations}\label{nominal}

Theoretical predictions of cross sections, reaction rates and other calculated quantities are strongly dependent on the type of nuclear models considered, as well as their associated parameters. This has been presented in a number of publications, regarding for instance the photon strength functions ~\cite{goriely2018a,goriely2018b}, level densities~\cite{PhysRevC.78.064307}, mass models~\cite{goriely2009}, optical models~\cite{bauge2004,goriely2023} or fission models~\cite{GORIELY2013115}.\\
A selection of nine specific nuclear models is made here to provide a fiducial default TALYS calculation.
These nominal nuclear model correspond to a compromise between advanced physics modelling and agreement with nuclear observables of relevance, including measured MACS and cross sections. They are listed below (values in parenthesis corresponds to TALYS options):

\begin{itemize}
    \item E1 photon strength function: Gogny D1M Hartree-Fock-Bogoluybov (HFB) plus QRPA \cite{goriely2018a} ("{\it strength 8}"),
    \item Level density: Skyrme-Hartree-Fock-Bogoluybov plus combinatorial level densities \cite{PhysRevC.78.064307}, ("{\it ldmodel 5}"),
    \item Nucleon-Nucleus optical model potential: phenomenological Koning-Delaroche optical model potential \cite{KONING2003231}, ("{\it jlmomp n}" ), 
    \item M1 photon strength function: Gogny D1M HFB+QRPA \cite{goriely2018b}, ("{\it strengthM1 8}"),
    \item Collective enhancement of the level density: not considered, ("{\it colenhance n}"),
    \item Width fluctuation corrections: Moldauer model \cite{Moldauer1980}, ("{\it widthmode 1}"),
    \item Nuclear mass model: Skyrme-HFB mass model (HFB-24) \cite{Goriely13a}, ("{\it massmodel 2}"), 
    \item Alpha optical model: Double folding potential \cite{Demetriou02}, ("{\it alphaomp 5}"),
    \item Fission barriers model: Skyrme-HFB fission barriers \cite{Goriely09b} ("{\it fismodel 5}").
\end{itemize}

\begin{figure}[htbp]
\centerline{
\resizebox{1.0\columnwidth}{!}{\rotatebox{-0}{\includegraphics[trim=6 18.5cm 2cm 5cm, clip]{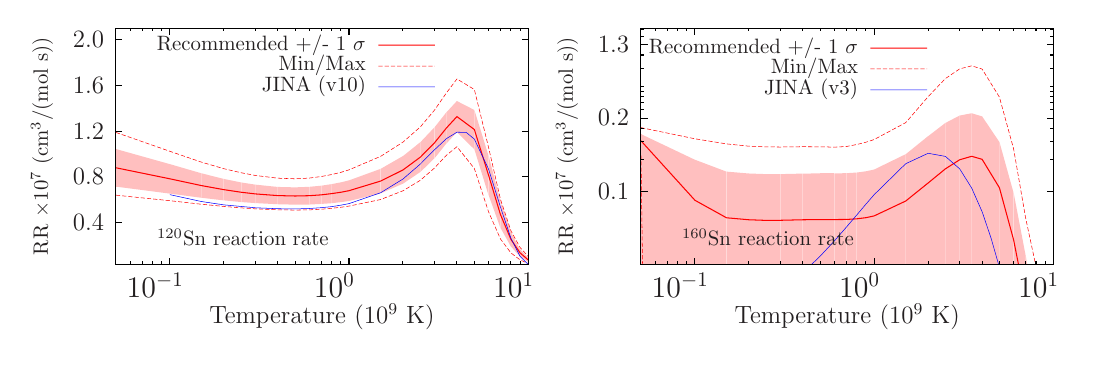}}}
}
\caption{Examples of recommended radiative neutron-capture rates for $^{120}$Sn (left) and $^{160}$Sn (right) from this work, compared to the JINA calculations~\cite{Cyburt_2010} (blue curves). The label ``RR'' indicates reaction rate. See text for details.
}
\label{sn120.rr} 
\end{figure}

\begin{figure}[htbp]
\centerline{
\resizebox{0.8\columnwidth}{!}{\rotatebox{-0}{\includegraphics[trim=2cm 11.5cm 2cm 9cm, clip]{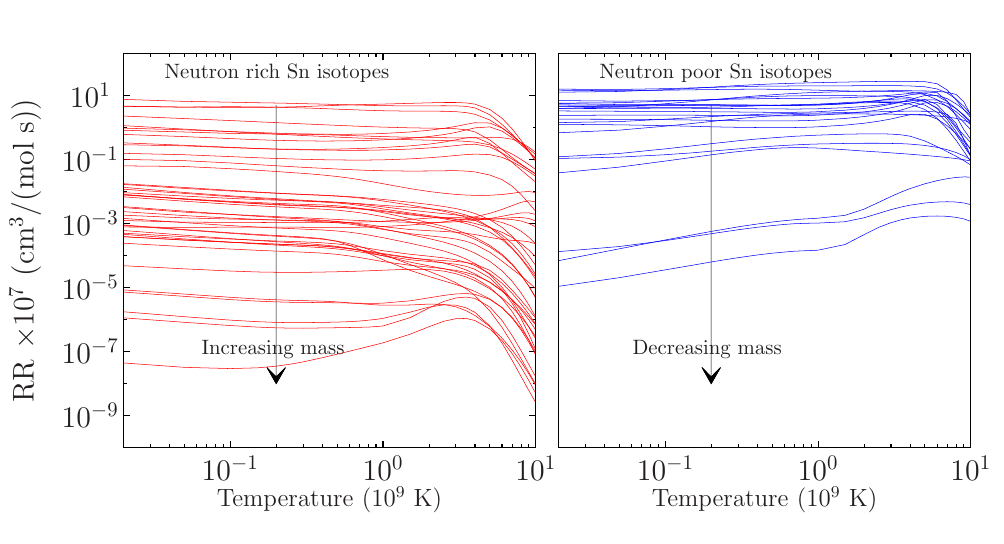}}}
}
\caption{Examples of recommended radiative neutron-capture rates for neutron-rich Sn isotopes (from $A=118$ to $A=161$, left panel) and neutron-poor Sn isotopes (from $A=97$ to $A=118$, right panel).
}
\label{sn.rr} 
\end{figure}

In a condensed manner, this selection of models is referred to as ``85n8n1255'', where each of these nine characters 
corresponds to a selected model from the previous list, following the same order. This selection of models is the one recommended for 
TALYS calculations for astrophysics applications (see section~\ref{verif} for comparison with experimental values). \\
The choice of the nuclear inputs to the cross section calculations is not motivated only by the comparison with known 
MACS or specific 
measured cross sections, but also directly with observables related to these inputs. For example, the choice of the Skyrme-HFB plus 
combinatorial nuclear level density model  \cite{PhysRevC.78.064307} is motivated by its (as much as possible) accurate description 
of low-lying schemes and $s$-wave resonance spacings. 
In addition, the selected models also describe in a satisfactory 
manner various other effects, such as the shell effects, pairing correlations, or rotational and vibrational enhancements. 
In the case of the photon strength function (Gogny D1M HFB+QRPA), the selected 
model predicts faily accurately photo-absorption data as well as resonance fluorescence data, average resonance capture data, 
average  $\Gamma_\gamma$ data (see Ref.~\cite{goriely2018b,Goriely19}). 
Naturally,  
other models can provide similar predictions, so that it remains difficult to favour one model over another. However, some models are 
systematically not reproducing specific observables and should consequently be avoided. For example the E1 generalized or standard 
Lorentzian models (GLO or SLO, respectively) photon strength functions (model "{\it strength}" 1 and 2 in TALYS) systematically 
under and overestimate, respectively, average radiative widths \cite{goriely2018b} and should better not be considered anymore.\\
In the following, the term ``recommended'' will be used instead of the nine characters ``85n8n1255''. Two examples for
$^{120}$Sn (stable) and $^{160}$Sn (unstable and very neutron rich) of recommended reaction rates are presented in Fig.~\ref{sn120.rr}.
The recommended values are indicated in plain red curves. The band (being one standard deviation, or 1$\sigma$), minimum as well as maximum values will be presented in section~\ref{variation}. The comparison with the calculations from Ref.~\cite{Cyburt_2010} indicates differences, more pronounced for the neutron-rich case. This is expected as the model uncertainties become more important when going further away from the experimentally known region, {\it i.e.} from the stability line. Similar trends are presented in Fig.~\ref{sn.rr} for considered neutron-rich and neutron-poor Sn isotopes (from mass 97 to mass 161).
Strong variations of reaction rates can be observed, spanning over many orders of magnitude. Neutron-rich isotopes tend to have a lower reaction rate (and capture cross section), mainly due to the decreasing neutron separation energies.\\ 
Tabulated data can be found in the online TENDL-astro 2023 database, under the section ``Recommended values'' for more than 8000 nuclides. Additional comparisons will be presented for other quantities in section~\ref{verif}, especially with selected experimental data. 

\subsection{Model variation}\label{variation}
\begin{figure}[htbp]
\centerline{
\resizebox{0.8\columnwidth}{!}{\rotatebox{-0}{\includegraphics[trim=1cm 11.5cm 3cm 2cm, clip]{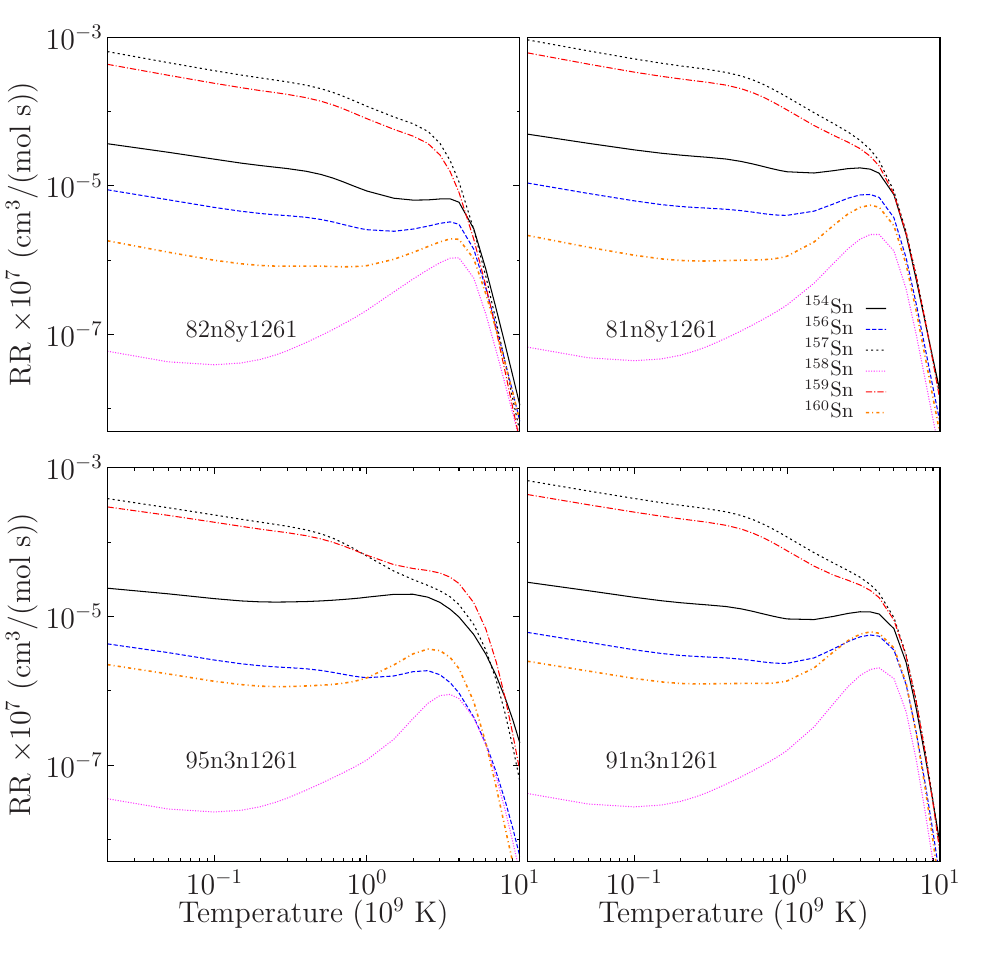}}}
}
\caption{Examples of radiative neutron-capture rates as a function of the temperature for neutron-rich Sn isotopes using four different model sets.
}
\label{sn.models.rr} 
\end{figure}

Once the  calculations for the recommended set of models have been performed, it is relatively straightforward to vary the choice of a number of models in order to assess their impact on the cross sections, reaction rates, partition functions and MACS. Such variations give an estimate of the impact of model uncertainties and the method to perform such variations is presented in the following. \\
A number of possibilities exist in the choice of these models, as defined in TALYS. In the case of the level density, there are six models to be selected (with the keyword {\it ldmodel} from 1 to 6), corresponding for instance to the Back-shifted Fermi gas Model \cite{Koning08}({\it ldmodel 2}), or the Skyrme-HFB plus statistical level densities \cite{Demetriou01} ({\it ldmodel 4}). Considering the TALYS numbering for the options, the following combinations of nuclear models are therefore applied, following the same order as in section~\ref{nominal}:
\begin{itemize}
    \item E1 photon strength function: "{\it strength}" 8 \cite{goriely2018a} or 9 \cite{Plujko18},
    \item Level density:  "{\it ldmodel}" 1 \cite{Koning08}, 2 \cite{Koning08} or 5 \cite{PhysRevC.78.064307},
    \item Optical model: "{\it jlmomp}" n \cite{KONING2003231} or y \cite{bauge2001} (for no or yes), 
    \item M1 photon strength function: "{\it strengthM1}" 3 \cite{Goriely18b} or 8 \cite{goriely2018b},
    \item Collective enhancement of the level density:  "{\it colenhance}"  n or y \cite{Koning08},
    \item Width fluctuation corrections: "{\it widthmode}" 0 (no correction), 1 \cite{Hofmann80} or 2 \cite{Moldauer1980},
    \item Nuclear masses model:  "{\it massmodel}" 0 \cite{Duflo95}, 1 \cite{Moller16}, 2 \cite{Goriely16a} or 3 \cite{goriely2009}, 
    \item Alpha optical model: "{\it alphaomp}" 5 \cite{Demetriou02} or 6 \cite{Avrigeanu2014},
    \item Fission barriers model: "{\it fismodel}" 1 \cite{Capote09}or 5 \cite{Goriely09b}.
\end{itemize}
All together, the number of combinations of the 9 model types is rather large, being $2^6\times 3^2\times 4=2304$ for fissile  and half (1152) for non-fissile nuclides. Among these possibilities, some are more trusted than others, and a number of them are undesirable; the following classification (as well as limitation) is therefore applied:
\begin{itemize}
    \item[A.] Most trusted combination (nominal calculation): recommended set (85n8n1255),
    \item[B.] Ten most trusted sets, based on the validation of the individual models \cite{talys} (as presented in section~\ref{nominal}): 91n3n1255, 85n8n1255, 
     92n3y1255, 92n3n1255, 91n3y1255, 95n3n1255, 82n8y1255, 82n8n1255, 81n8y1255, 81n8n1255. For these models, the last four models are 
     fixed to the recommended values, as well as the choice of the optical model (third digit). Specific combinations of level density 
models with photon strength functions (both E1 and M1) are also selected. 
    \item[C.] For the other cases, 480 model combinations are considered for non-fissile nuclides (masses below 210), and the double (960) for fissile ones (due to the variation of the model for fission barriers).
\end{itemize}
In the case of the group C, not all combinations are considered. For instance, the choice of the E1 and M1 photon strength functions are coherently linked: if four possibilities exist (2 models for each function), only two are considered:  the D1M HFB+QRPA model \cite{goriely2018a,goriely2018b} ``{\it strength=8}'' with ``{\it strengthM1=8}'' or the Simple Modified Lorentzian (SMLO) model \cite{Plujko18,Goriely18b} ``{\it strength=9}'' with ``{\it strengthM1=3}''. This allows to keep a theoretical consistency between E1 and M1 photon strength functions. Additionally, in the case of the Skyrme HFB plus combinatorial level densities \cite{PhysRevC.78.064307} (``{\it ldmodel=5}''), the collective enhancements of the level density are never selected (``{\it colenhance=n}'') since it is already embedded in the numerical tables.  These selections are thus limiting the number of possibilities for the model combinations and lead to a total of 960 possibilities for fissile nuclides (with $A\geq 210$) and 480 for the other ones. \\
An example of calculated reaction rates is presented in Fig.~\ref{sn120.rr} for $^{120}$Sn and $^{160}$Sn. The uncertainty bands and the extreme values are obtained for the case C, based on the 480 considered models. One can observe larger differences (larger extreme values and uncertainties) for $^{160}$Sn, far from the stability line. The comparison with the JINA calculations~\cite{Cyburt_2010} does not show unrealistic differences. This example indicates the importance of considering different reaction models, especially for short-lived exotic nuclides for which experimental information is scarce or non-existing. To emphasize this effect, the reaction rates for very neutron-rich nuclides of Sn are presented in Fig.~\ref{sn.models.rr} for four different model sets.
The four selected model sets are from the case B ({\it i.e.} among the 10 most trusted model sets). One can observe differences of one order of magnitudes for a selection of models and nuclides. The impact of the level density models (two phenomenological models and one microscopic model, if one compares plots on the same rows of Fig.~\ref{sn.models.rr}) is particularly pronounced, and to a less extent for the photon strength functions (comparing top and bottom right plots). \\
The TENDL-astro 2023 database includes the variations due to different models for about 8000 nuclides, for all mentioned quantities of interest. Additionally, calculated quantities (with averages, standard deviations and extreme values) are proposed for the cases B and C (for case A, only the values from the recommended model are presented, followed by the uncertainties from the case B, {\it i.e.} from the variations of the 10 most trusted model sets).\\
In the next sections, comparisons for the MACS and neutron-capture cross sections at 30~keV are presented, including  experimental data when they exist.

\section{Comparisons with selected quantities}\label{verif}
Due to the large amount of calculated quantities, not all possibilities can be presented and a limited number of comparisons is performed. 
It is also not practical to overload the reader with many statistical indicators, so that we will simply introduce here 
two root-mean-square (rms) quantities $\epsilon_{\rm rms}$ and $f_{\rm rms}$ to represent the results, as detailed below.

\subsection{Statistical indicators}
The comparison of the different cross sections and reaction rates can be cumbersome, and it is useful to use simple and representative statistical quantities. Apart from the traditional mean, median and standard deviation, the following two indicators for the dispersion of a  population are used (Eqs.~(\ref{erms}) and (\ref{frms})).
\begin{eqnarray}
r_i &=& \frac{Q_i}{Q_{\rm ref}}\nonumber\\
\epsilon_{\rm rms} &=& \exp\left(\frac{1}{N}\sum_{i=1}^{N}\ln(r_i)\right)\label{erms}\\
f_{\rm rms} &=& \exp\sqrt{\frac{1}{N}\sum_{i=1}^{N}\ln^2(r_i)}\label{frms}
\end{eqnarray}
In these equations, $Q_i$ is either a calculated cross section, reaction rate or MACS, where $i$ indicates the considered model set (as presented 
in sections~\ref{nominal} and \ref{variation}). The quantity $Q_{\rm ref}$ is similar to $Q_i$, but for a reference choice, {\it i.e.} either the 
recommended model set (``85n8n1255''), or the selected experimental data in the case of the MACS. The $f_{\rm rms}$ quantifies the accuracy of 
the predictions, it is a quantity always larger (or equal) than 1, and the deviation from 1 indicates on average ``{\it how far}'' the prediction from 
the reference is. For example, a value of 1.40 indicates that the prediction is on average 40~\% with respect to the reference. 
The second quantity, $\epsilon_{\rm rms}$, can be smaller than 1 and indicates if a prediction overpredicts ($\epsilon_{\rm rms}>1$) or 
underpredicts ($\epsilon_{\rm rms}<1$) the reference set.

\subsection{Comparisons for MACS values at 30 keV}
\begin{figure}[htbp]
\centerline{
\resizebox{0.8\columnwidth}{!}{\rotatebox{-0}{\includegraphics[trim=0cm 12cm 8cm 12cm, clip]{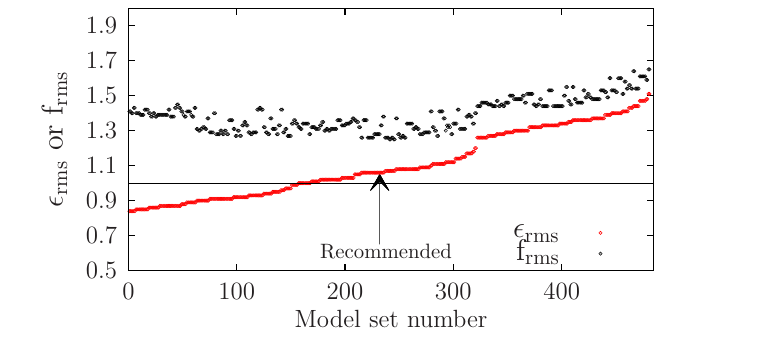}}}
}
\caption{$\epsilon_{\rm rms}$ and $f_{\rm rms}$ for (n,$\gamma$) MACS  at 30 keV, for 480 models. Ratios are the model $i$ over the experimental MACS, as 
compiled in the KADoNiS database \cite{dillmann2006,DILLMANN2014171}, for $40 \le A \le 210$. }
\label{epsilonMACS} 
\end{figure}

\begin{figure}[htbp]
\centerline{
\resizebox{0.8\columnwidth}{!}{\rotatebox{-0}{\includegraphics[trim=0cm 12cm 8.5cm 11.5cm, clip]{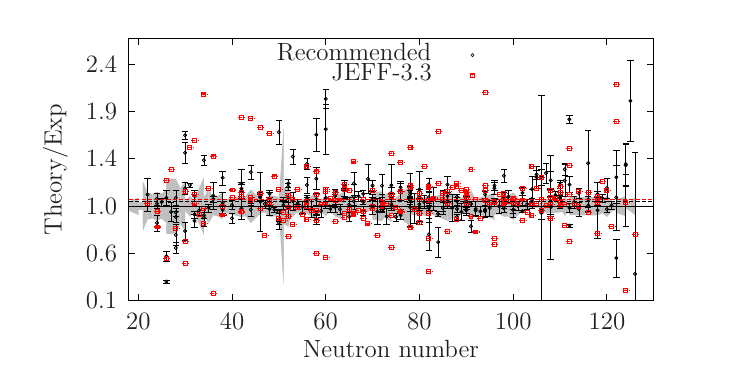}}}
}
\centerline{
\resizebox{0.8\columnwidth}{!}{\rotatebox{-0}{\includegraphics[trim=0cm 12cm 8.5cm 11.5cm, clip]{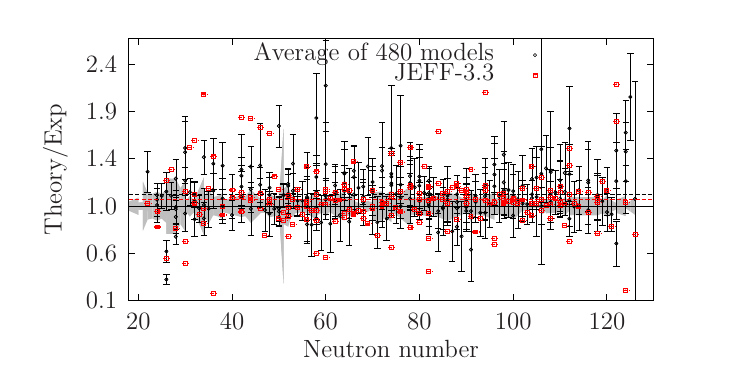}}}
}
\caption{Top: Ratio of the calculated (n,$\gamma$) MACS (either recommended (black circles) or JEFF-3.3 (red squares), being the quantity Theory) over the KADoNiS values (being the quantity Exp) for neutron numbers 
corresponding to $40 \le A \le 210$.
Uncertainties in black vertical lines correspond to the standard deviations of the 10 most trusted model sets (Group B), and gray bands correspond to the experimental uncertainties. Dashed lines are median values and the full line indicates Theory/Exp.=1. Bottom: same with C being the average of 480 model variations (Group C).
}
\label{ratioMACS} 
\end{figure}

One observable of particular relevance for astrophysical applications is the MACS at 30 keV. It has been evaluated based on measurements and reported in 
the KADoNiS database~\cite{dillmann2006}. Many comparisons between this database and TALYS calculations was already 
presented~\cite{goriely2008,fleming2015,PhysRevC.90.034619,KONING20191}, mainly considering either the default model set, or studying 
the effect of a particular model. To present a systematic study of the model variation with respect to the KADoNiS data, Fig.~\ref{epsilonMACS} 
presents both the $f_{\rm rms}$ and the $\epsilon_{\rm rms}$ for the 480 considered variations for non-fissile nuclides. The KADoNiS data used 
in the comparison correspond to $40 \le A \le 210$ nuclides to avoid non-statistical effects for light species and  uncertainties associated with fission probabilities for heavier nuclei.
%
In this figure, the reference data are the MACS at 30 keV from the KADoNiS database. The order of the model sets on the x-axis has been simply ordered from 
the lowest $\epsilon_{\rm rms}$ value to the largest one. The recommended model set leads to $\epsilon_{\rm rms}=1.06$ and 
$f_{\rm rms}=1.28$, as indicated in Fig.~\ref{epsilonMACS}
(the recommended model has not been explicitly tuned to fit the measured MACS). 
Different values, in better agreement with the KADoNiS results, would be obtained otherwise.\\
Other model combinations have similar $f_{\rm rms}$ values, and possibly better $\epsilon_{\rm rms}$, but the recommended model set still provides one of the best $f_{\rm rms}$ values. For comparison, the JEFF-3.3, JENDL-5.0, ENDF/B-VIII.0 and TENDL-2023 libraries~\cite{plompen2020,iwamoto2023,BROWN20181,KONING20191}, providing dedicated evaluated cross sections for single isotopes, give rise to values of  $\epsilon_{\rm rms}=1.32, 
1.15, 1.17$ and $0.88$, respectively (the corresponding $f_{rms}$ values are 2.8, 2.0, 2.1 and 1.4). 
As mentioned, the selection of the models for the recommended set is also based on the comparison with other experimental data, as well as with the expected predictive power far from the stability line in relation to the sound physics included in the respective nuclear models. \\
In terms of C/E values for each nuclide, cases corresponding to the recommended model set, the average of the 480 
model sets and the JEFF-3.3 library are presented in Fig.~\ref{ratioMACS}. For the recommended set, the uncertainties on the 
C/E values come from the spread of the 10 most trusted model combinations (being one standard deviation).
As for the previous comparison, only nuclides with a mass between 40 and 210 are plotted. The dotted red and black lines correspond to the average C/E values, considering the different calculated cases.
One can also observe the increase of the calculated uncertainties (due to the variations of either 10 or 480 models) between the two figures. \\
These comparisons with the MACS from the KADoNiS database globally indicate a good prediction power of the recommended model set. Uncertainties due to the model 
variation are not negligible, indicating a potential impact on astrophysical simulations.  

\subsection{Comparisons for neutron capture cross sections at 30 keV}
\begin{figure}[htbp]
\centerline{
\resizebox{0.7\columnwidth}{!}{\rotatebox{-0}{\includegraphics[trim=0cm 20cm 8.5cm 3cm, clip]{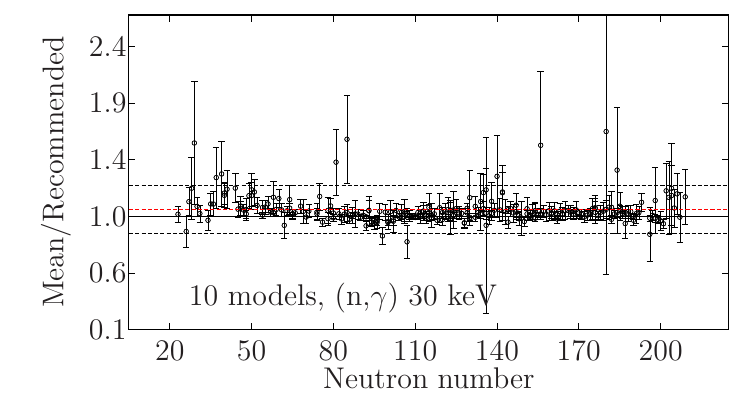}}}
}
\centerline{
\resizebox{0.7\columnwidth}{!}{\rotatebox{-0}{\includegraphics[trim=0cm 20cm 8.5cm 3cm, clip]{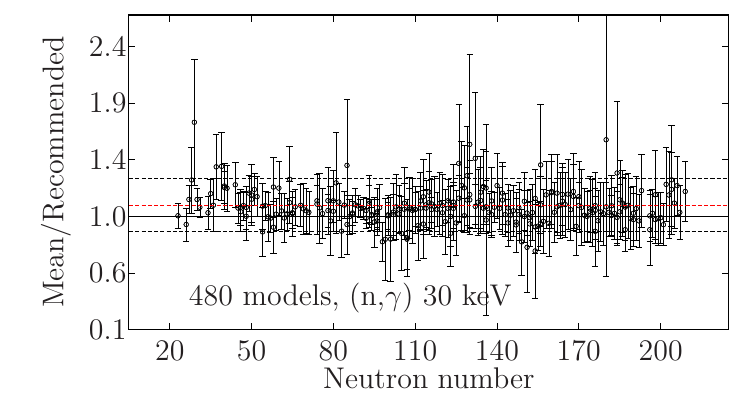}}}
}
\caption{Ratios of averages of 10 (top) and 480 (bottom) models over the  recommended TALYS model set for the radiative neutron capture cross section at 30 keV. See text for details. Only ratios for the 250 nuclei with known resonance properties (``adjust\_w'' group) are shown.} 
\label{ratio.10models} 
\end{figure}

\begin{figure}[htbp]
\centerline{
\resizebox{1.0\columnwidth}{!}{\rotatebox{-0}{\includegraphics[trim=0 9cm 0 6cm, clip]{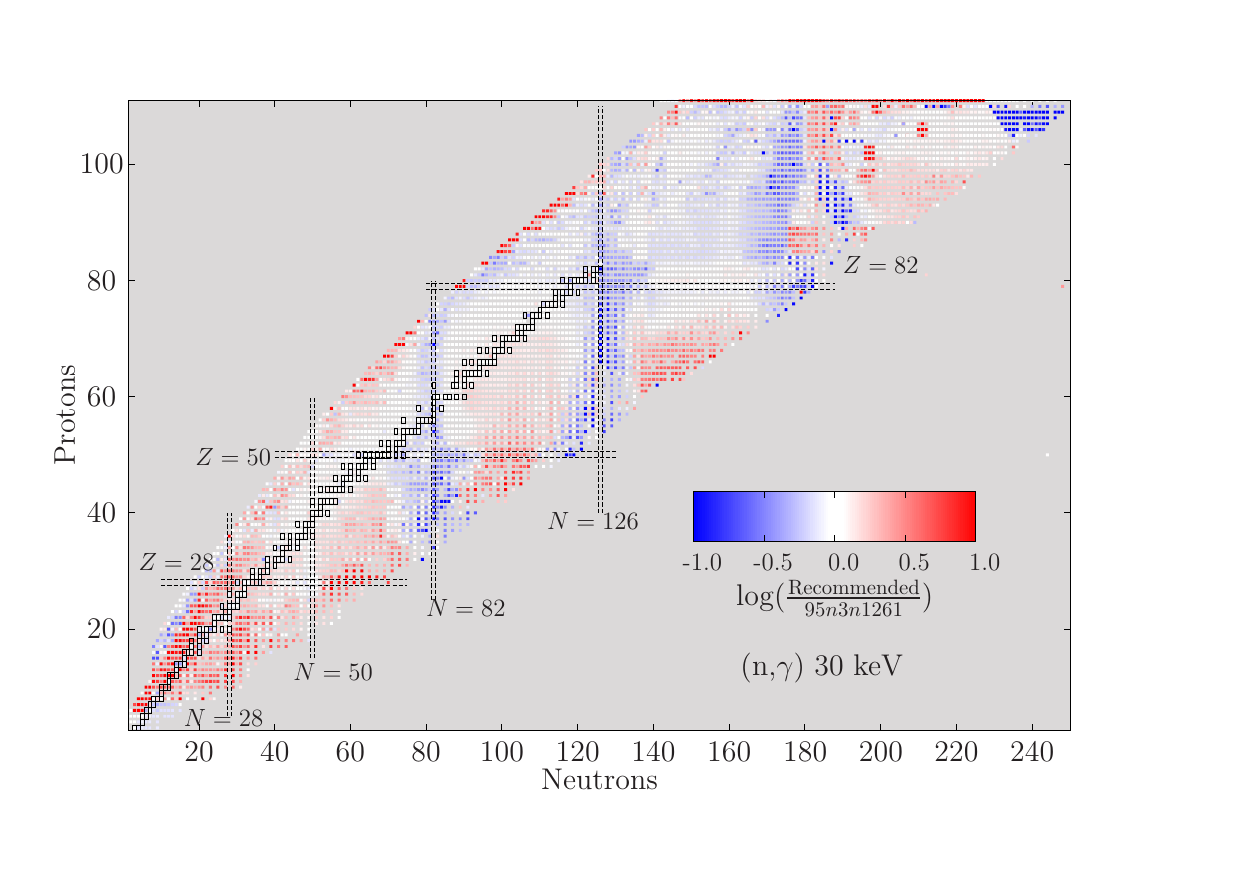}}}
}
\caption{Illustration in the ($N$,$Z$) plane of the ratios between the recommended model set and the model 95n3n1261 for the radiative neutron-capture cross sections at 30 keV. The color scale is logarithmic, indicating large range of variations. Square boxes indicates the location of stable nuclides.}
\label{81n8n12.91n3n12.30keV} 
\end{figure}

In this section, neutron capture cross sections at 30 keV are compared between the recommended model sets and other ones. 
As for the MACS, the agreement with the average 
of 10, 480 or 960 models is expected to indicate higher deviations if more model variations are included. Additionally, 
as a number of models are in relative agreement for stable nuclides, such deviations are also expected to increase when 
going further away from the stability line. The considered nuclides are therefore divided in three sub-groups: one for 
the nuclides with known resonance data \cite{Capote09} (which have been used to fine-tune photon strength functions and 
level densities) (called ``known\_res''), one including nuclides with measured masses (excluding the 
previous  ``known\_res'' group, and called ``known\_m''), and finally one for the remaining nuclides (called ``others''). 
These groups contain, respectively about 250, 2700 and 1700 for non-fissile nuclides, and 20, 550 and 3000 for the fissile 
ones. Table~\ref{table_stat1} gives the $f_{\rm rms}$ values for the case of 10 models (the $r_i$ of Eq.~(\ref{frms}) 
corresponds to $Q_{\rm 10 models}/Q_{\rm ref}$), only for non-fissile nuclides (similar trends are obtained for other 
quantities, such  as $\epsilon_{\rm rms}$,  for fissile nuclides, or for 480 models instead of 10).
\begin{table}[htbp]
\caption{$f_{\rm rms}$ values for the 10 model sets with respect to the recommended model set) for 
$5<A<220$ nuclei. ``XS'' means cross sections, and ``RR'' reaction rates. See text for the explanation of the nuclide groups. The (n,$\gamma$), (n,p) and (n,$\alpha$) XS are calculated at 30~keV, 10~MeV and 10~MeV, respectively and the corresponding RR at $T=0.3$, 10 and 10~GK.
}
\begin{center}
{
\begin{tabular}{l cc || cc || cc}
\hline
 Nuclide group&   Quantity &      $f_{\rm rms}$ &   Quantity &       $f_{\rm rms}$ &   Quantity &       $f_{\rm rms}$\\
\hline
          & \multicolumn{2}{c||}{(n,$\gamma$)}  & \multicolumn{2}{c||}{(n,p)} & \multicolumn{2}{c}{(n,$\alpha$)}    \\
known\_res & XS 30 keV                &   1.14 & XS 10 MeV       & 1.29    & XS 10 MeV       & 1.69 \\
known\_m  &           "              &   1.81 &        "        & 1.74    &        "        & 2.82 \\  
others    &           "              &   2.07 &        "        & 2.28    &        "        & 15.0 \\
\hline
          & \multicolumn{2}{c||}{(n,$\gamma$)}  & \multicolumn{2}{c||}{(n,p)} & \multicolumn{2}{c}{(n,$\alpha$)}    \\
known\_res & RR 0.3 GK                 &   1.13 & RR 10 GK        &  1.91 & RR 10 GK   & 1.83 \\
known\_m  &        "                  &   1.80 &       "         &  2.09 &     "      & 2.44 \\ 
others    &        "                  &   2.00 &       "         &  3.80 &     "      & 5.13 \\    
\hline
\hline
\end{tabular}
}
\end{center}
\label{table_stat1}
\end{table}
As observed, the $f_{\rm rms}$ values increase for nuclides further away from the stability line, not only for cross sections but also for reaction rates. In the case of proton- or alpha-induced reactions, similar trends exist (not presented in the table). 
Local ratios for nuclides from the adjust\_w group, between the average of 10 (or 480) model sets and the recommended set is presented in Fig.~\ref{ratio.10models}. 
One can observe that the calculated uncertainties (vertical error bars) in the figure are naturally increasing from 10 to 480 model sets, being on the average 7.5~\% and 20~\%. 
The averages (and standard deviations, presented as dotted lines in Fig.~\ref{ratio.10models}) of the ratios are 1.06 ($\pm 0.19$) for 10 model sets and 1.09 ($\pm 0.21$) for 480 model sets.  For these nuclides, the differences are not strongly pronounced, as there characteristics are experimentally well-known, leading to limited variations between models. \\
Differences between models are expected to increase for less stable nuclides, exhibiting patterns following model properties. An example of such variations are presented in Fig.~\ref{81n8n12.91n3n12.30keV}.
The ratio of the neutron capture cross sections at 30 keV, for all considered nuclides, between the recommended model set and the combination 95n3n1261 is plotted using a logarithmic scale.
Local patterns can be observed, globally indicating strong variations for the majority of nuclides away from the stability line. 
Specific effects in the vicinity of the closed neutron shells $N=82$ and $N=126$ can be observed, with a strong 
decrease of the MACS ratios. This impact is due to the different way of treating shell effects in masses, nuclear level densities 
and photon strength functions in the respective models considered.
It should be noted that the values presented in this figure correspond the state of knowledge from a selection of models (as included in TALYS). Another selection, for instance including
the direct capture mechanism, will lead to different ratios.
Such a mapping, for this specific model set, as well as for other ones, shows the importance of 
taking into account model variations for this quantity. Similar variations were observed for reaction rates and MACS, 
as already presented in Refs.~\cite{rochman2023,Goriely2023iop,goriely2023}.

\section{Conclusion}
The impact of nuclear reaction models in the case of unstable nuclides in the astrophysics context should not be neglected. 
The present work confirms that reaction models can lead to very different quantities of interest. In order to propose such 
results for more than 8000 nuclides to potential users, the TENDL-astro database, version 2023, is now available online, 
together with the TENDL libraries (https://tendl.web.psi.ch/tendl\_2023/astro/astro.html). Results include cross sections 
for neutron-, proton- and $\alpha$-induced cross sections and reaction rates, Maxwellian averaged cross sections (at 
30~keV), as well as partition functions. It is based on TALYS calculations, with a selection of model sets, varying 
all together 9 different reaction models, including level densities, photon strength functions, and optical potentials. 
The different model combinations (or model sets) encompass 480 cases for non-fissile nuclides, and 960 for the fissiles 
ones. With these variations, a recommended model set is proposed, together with 10 preferred  model sets. The database 
also proposes averages, standard deviations and other statistical indicators. It is likely to evolve in the future, 
as TALYS modelling is continuously improved, and other quantities of interest might be included.

\section*{Acknowledgements}
This work was partly funded by the European Union’s Horizon 2020 Research and Innovation
Programme under grant agreement No 847552  (project SANDA, Supplying Accurate Nuclear Data for energy and non-energy Applications). This work was supported by the Fonds de la Recherche Scientifique (F.R.S.-FNRS) and the Fonds Wetenschappelijk Onderzoek - Vlaanderen (FWO) under the EOS Projects nr O022818F and O000422F and by the European Union (ChECTEC-INFRA, project no. 101008324). 

\newpage
\bibliographystyle{ieeetr}
\bibliography{bibliography.bib}

\end{document}